\documentclass[12pt]{iopart}
\expandafter\let\csname equation*\endcsname\relax
\expandafter\let\csname endequation*\endcsname\relax
\usepackage{amsmath}
 
\begin{document}

\title{Novel distributional Laplacians and a Coulomb-gauge problem}
\author{V Hnizdo$^1$ and G Vaman$^2$} 
\address{$^1$ 2044 Georgian Lane, Morgantown, WV 26508, USA}
\address{$^2$ Aleea Callatis 1, Bucharest, Romania}
\eads{\mailto{hnizdo2044@gmail.com} and \mailto{getavaman@gmail.com}}
\begin{abstract}
Novel distributional Laplacians of an arsinh function and a related logarithm function are derived. The method used to establish the former result is employed in a calculation of the gauge transformation function $\chi_{\rm C}$ for the case of a uniformly moving point charge.
\end{abstract}
%\noindent{\it Keywords\/}: classical electrodynamics, Lorenz gauge, Coulomb gauge, gauge function

\section{Introduction}
The classical Laplacian of the function ${\rm arsinh}[a(z -b)/s]$, where $s=(x^2+y^2)^{1/2}$ and $z$ are cylindrical coordinates, and $a$ and $b$ are arbitrary independent constants, is given by\footnote{We use the symbol $\Delta$ instead of $\nabla^2$ for the Laplacian operator  so that its distributional inverse, which we will introduce in this paper, can be denoted simply and legibly.}
\begin{align}
\Delta\,{\rm arsinh}\frac{a(z -b)}{s}=\frac{a(1-a^2)(z-b)}{[s^2+a^2(z-b)^2]^{3/2}}.
\label{classL}
\end{align}
It may come as a surprise that the distributional (generalized) Laplacian adds to the classical one a delta-function term,
\begin{align}
\bar{\Delta}\, {\rm arsinh}\frac{a(z -b)}{s}=\frac{a(1-a^2)(z-b)}{[s^2+a^2(z-b)^2]^{3/2}}
-{\rm sgn}[a(z-b)]\frac{\delta(s)}{s},
\label{distrL}
\end{align}
where the bar in $\bar{\Delta}$ denotes it as a distributional operator, ${\rm sgn}(\cdot)$ is the sign function and the delta function $\delta(s)$ is normalized as $\int_0^{\infty}\rmd s\, \delta(s)=1$.
Several novel distributional Laplacians involving the arsinh function can be obtained with specific values of the constants in Eq.\,(\ref{distrL}). For example, when $a=1$ and $b=0$, we obtain 
\begin{align}
\bar{\Delta}\,
 {\rm arsinh}\frac{z}{s}=-{\rm sgn}(z)\frac{\delta(s)}{s},
\label{ash(z/s)}
\end{align}
which is similar to the distributional  Laplacian of the logarithm function, 
\begin{align}
\bar{\Delta}\, \ln (s)=\frac{\delta(s)}{s},
\label{ln(s)}
\end{align}
an informal proof of which was given in \cite{RH}.
%\footnote{Interestingly, 
%$\bar{\Delta}\ln (r)=-2\pi\delta({\bi r})$ in 2D (\cite{Kan}, p 274).} 
The generation of an additional  delta-function term by the application of the distributional Laplacian in cylindrical coordinates to the function ${\rm arsinh}[a(z-b)/s]$  is similar to the distributional Laplacian in spherical coordinates 
generating  additional distributional terms when it is applied to some functions \cite{Cant,Etx}.

In Section 2, we give an informal proof of  the distributional Laplacian (\ref{distrL}); another, related, distributional Laplacian is there derived also. 
In Section 3, the method used to establish the mathematical result (\ref{distrL})  is employed in a calculation of the gauge function  of the transformation of the Lorenz-gauge potentials of a uniformly moving charge to the Coulomb gauge. The last section contains some concluding remarks.

\section{Informal proofs} 

The distributional relation (\ref{distrL}) can be established by solving  the (distributional) Poisson equation
\begin{align}
\bar{\Delta}\, 
 g(s,z)=\frac{a(1-a^2)(z-b)}{[s^2+a^2(z-b)^2]^{3/2}}
 -{\rm sgn}[a(z-b)]\frac{\delta(s)}{s},
\label{Peq}
\end{align}
using the standard integral representation of its solution, the operator of which we term the inverse Laplacian and denote by $\bar{\Delta}^{-1}$. Thus
\begin{align}
g(s,z)&=\bar{\Delta}^{-1}\left[\frac{a(1-a^2)(z-b)}{[s^2+a^2(z-b)^2]^{3/2}}
-{\rm sgn}[a(z-b)]\frac{\delta(s)}{s}\right] \nonumber \\
&\equiv -\frac{1}{4\pi}\int \frac{\rmd^3 r'}{|{\bi r}-{\bi r}'|}\left[\frac{a(1-a^2)(z'-b)}
{[s'^2+a^2(z'-b)^2]^{3/2}}-{\rm sgn}[a(z'-b)]\frac{\delta(s')}{s'}\right].
\label{solvP}
\end{align}
Since we use cylindrical coordinates, we employ  in (\ref{solvP}) a cylindrical-coordinate expansion of the inverse distance (\cite{Jack}, p 140),
\begin{align}
\frac{1}{|{\bi r}-{\bi r}'|} =\sum_{m=-\infty}^{\infty}\int_0^{\infty}\rmd k\,
{\rm e}^{{\rm i}m(\phi-\phi')} J_m(k s)J_m(k s'){\rm e}^{-k(z_>-z_<)},
\label{cylexp}
\end{align}
where $J_m(\cdot)$ are the Bessel functions of the first kind of order $m$ and $z_>$ ($z_<$) is the greater (lesser) of the $z$-coordinates of the vectors $\bi r$ and ${\bi r}'$.

We calculate first the inverse Laplacian of the non-delta-function part of the RHS of (\ref{Peq}),  
\begin{align}
&\bar{\Delta}^{-1} \frac{a(1-a^2)(z-b)}{[s^2+a^2(z-b)^2]^{3/2}} \nonumber \\
&\quad=\frac{a(a^2-1)}{2}\int_0^{\infty}\rmd k\,J_0(k s)
\int_{-\infty}^{\infty}\rmd z'(z'-b){\rm e}^{-k(z_>-z_<)}\int_0^{\infty}s'\,\rmd s'
\frac{J_0(k s')}{[s'^2+a^2(z'-b)^2]^{3/2}} \nonumber \\
&\quad=\frac{a(a^2-1)}{2|a|}\int_0^{\infty}\rmd k\,J_0(k s) 
\int_{-\infty}^{\infty}\rmd z'{\rm sgn}(z'-b){\rm e}^{-k(z_>-z_<)}\,{\rm e}^{-|a(z'-b)|k},
\label{nondlt1}
\end{align}
where the 2nd line is  simplified as a result of $\int_0^{2\pi}\rmd \phi'\exp[{\rm i}m(\phi-\phi')]=2\pi\delta_{m0}$ and the 3rd line employs the result
\begin{align}
\int_0^{\infty}s'\rmd s'\frac{J_0(k s')}{[s'^2{+}a^2(z'{-}b)^2]^{3/2}}
=\frac{1}{|a(z'-b)|}{\rm e}^{-|a(z'-b)|k},
\label{Prud2}
\end{align}
obtained using  a tabulated integral (\cite{Prud}, item 2.12.4(28)). The evaluation of the integral with respect to $z'$ in (\ref{nondlt1}) is cumbersome, but straightforward. It yields 
\begin{align}
&\int_{-\infty}^{\infty}\rmd z'{\rm sgn}(z'-b){\rm e}^{-k(z_>-z_<)}\,{\rm e}^{-|a(z'-b)|k}
\nonumber \\
&\quad=\int_{-\infty}^{z}\rmd z'{\rm sgn}(z'-b){\rm e}^{-k(z-z')}\,{\rm e}^{-|a(z'-b)|k}
+\int_{z}^{\infty}\rmd z'{\rm sgn}(z'-b){\rm e}^{-k(z'-z)}\,{\rm e}^{-|a(z'-b)|k} \nonumber \\
&\quad=\frac{2}{(a^2-1)k}\bigg\{
\begin{array}{cc}
1-{\rm e}^{-|a|(z-b)k}-(1-{\rm e}^{-(z-b)k}),& z>b\\ 
1-{\rm e}^{-(b-z)k}-(1-{\rm e}^{-|a|(b-z)k}),& z<b
\end{array}\bigg.
\label{nondlt2}
\end{align}  
The peculiar writing 
of this result is to facilitate the use of a tabulated integral (\cite{Prud}, item 2.12.8(5)),
\begin{align}
\int_0^{\infty} \rmd k\, \frac{1}{k}(1-{\rm e}^{-\rho k})J_0( k s)= {\rm arsinh}\,\frac{\rho}{s},
\label{Prud}
\end{align}
in the remaining integration in (\ref{nondlt1}), which gives
\begin{align}
\bar{\Delta}^{-1} \frac{a(1-a^2)(z-b)}{[s^2+a^2(z-b)^2]^{3/2}}
&={\rm sgn}(a)\left[{\rm arsinh}\frac{|a|(z-b)}{s}-{\rm arsinh}\frac{z-b}{s}\right]\nonumber \\
&={\rm arsinh}\frac{a(z-b)}{s}-{\rm sgn}(a)\,{\rm arsinh}\frac{z-b}{s}.
\label{invLnondlt}
\end{align}
Since ${\rm arsinh}(-x)=-{\rm arsinh}(x)$, this result holds true for both $z>b$ and $z<b$. 

We now calculate the inverse Laplacian of the delta-function part of (\ref{Peq}), 
\begin{align}
&\bar{\Delta}^{-1}\left[-{\rm sgn}[a(z-b)]\frac{\delta(s)}{s}\right]\nonumber \\
&\quad\quad=\frac{{\rm sgn}(a)}{2}\int_0^{\infty}\rmd k\,
J_0(k s)\int_{-\infty}^{\infty}\rmd z'\, {\rm sgn}(z'-b){\rm e}^{-k(z_>-z_<)}
\nonumber\\
&\quad\quad=\frac{{\rm sgn}(a)}{2}\int_0^{\infty}\rmd k\,
J_0(k s)\left\{{\rm e}^{-k z}\left[-\int_{-\infty}^b \rmd z'{\rm e}^{k z'}+\int_b^z\,\rmd z'{\rm e}^{kz'}\right]+{\rm e}^{k z}\int_z^{\infty}\rmd z'\,{\rm e}^{-k z'}\right\} \nonumber \\
&\quad\quad ={\rm sgn}(a)\int_0^{\infty}\rmd k\,J_0(k s)\frac{1}{k} [1- {\rm e}^{-k(z-b)}], 
\label{deltapt1}
\end{align}
where the 2nd line is simplified as a result of the integral  with respect to $\phi'$ equaling  
$2\pi\delta_{m0}$
 and of the integral with respect to $s'$ yielding a unity factor  $J_0(0)$ on account of the factor $\delta(s')/s'$; in the 3rd line, $z>b$ is assumed.
The last line of Eq.\,(\ref{deltapt1}) 
can be evaluated using   the integral (\ref{Prud}), yielding 
\begin{align}
\bar{\Delta}^{-1}\left[-{\rm sgn}[a(z-b)]\frac{\delta(s)}{s}\right]={\rm sgn}(a)\,{\rm arsinh}\frac{z-b}{s},
\label{invLdlt}
\end{align}
which, as the result (\ref{invLnondlt}),  holds true for both $z>b$ and $z<b$.

Adding  now Eqs.\,(\ref{invLnondlt}) and (\ref{invLdlt}) gives
\begin{align}
\bar{\Delta}^{-1}\left[ \frac{a(1-a^2)(z-b)}{[s^2+a^2(z-b)^2]^{3/2}}-{\rm sgn}[a(z-b)]\frac{\delta(s)}{s}\right]={\rm arsinh}\frac{a(z-b)}{s}. 
\end{align}
The Poisson equation (\ref{Peq}) is thus solved by
\begin{align}
g(s,z)={\rm arsinh}\frac{a(z-b)}{s},
\end{align}
which establishes the distributional-Laplacian relation (\ref{distrL})
as holding true. 

Using the identity ${\rm arsinh}(x)=\ln(x+\sqrt{1+x^2})$ and relation  (\ref{ln(s)}) for $\bar{\Delta}
\ln(s)$, the 
distributional relation (\ref{distrL}) can be written as
\begin{align}
\bar{\Delta}\ln[a(z{-}b)+\sqrt{s^2+a^2(z{-}b)^2}] =\frac{a(1-a^2)(z-b)}{[s^2+a^2(z-b)^2]^{3/2}}+\big[1-{\rm sgn}[a(z{-}b)]\big]\frac{\delta(s)}{s}.
\label{LapLn}
\end{align}
Intriguingly, the distributional Laplacian of this logarithmic function generates a delta-function term 
only when $a(z-b)$ is negative. Unfortunately, this result cannot be established independently of the validity of (\ref{distrL}) and (\ref{ln(s)}) by a calculation of  the inverse Laplacian of the RHS of (\ref{LapLn})
 since such calculation  diverges due to the divergence of 
$\bar{\Delta}^{-1}[\delta(s)/s]$. 
But the  distributional Laplacian (\ref{LapLn}) 
can be established as
\begin{align}
\bar{\Delta}\ln[a(z{-}b)+\sqrt{s^2+a^2(z{-}b)^2}]={\rm w}\!\lim_{\epsilon\rightarrow 0}\Delta
\ln[a(z{-}b)+\sqrt{s^2+a^2(z{-}b)^2+\epsilon^2}],
\label{Deltawlim}
\end{align}
where the symbol wlim denotes the weak limit (see, e.g., \cite{Vlad}), 
since it can be shown that 
\begin{align}
&\lim_{\epsilon\rightarrow 0}\int_0^{\infty} s\,\rmd sf(s)\,
\Delta\ln[a(z-b)+\sqrt{s^2+a^2(z-b)^2+\epsilon^2}]
\nonumber \\
&=a(1-a^2)(z-b)\int_0^{\infty}\frac{s\,\rmd s f(s)}{[s^2+a^2(z-b)^2]^{3/2}}+\big[1-{\rm sgn}[a(z-b)]\big]f(0),
\label{limDelta}
\end{align}
where $f(s)$ is any well-behaved test function. An outline of this calculation is given in Appendix.

\section{Gauge transformation function $\chi_{\rm C}$}

The gauge function $\chi_{\rm C}({\bi r},t)$ that transforms the Lorentz-gauge potentials $\Phi_{\rm L}
({\bi r},t)$ and ${\bi A}_{\rm L}({\bi r},t)$  
to the Coulomb-gauge potentials $\Phi_{\rm C}({\bi r},t)$ and ${\bi A}_{\rm C}({\bi r},t)$ according to
\begin{align}
\Phi_{\rm C}({\bi r},t)=\Phi_{\rm L}({\bi r},t)-\frac{\partial\chi_{\rm C}({\bi r},t)}{c\partial t},
\quad {\bi A}_{\rm C}({\bi r},t)={\bi A}_{\rm L}({\bi r},t)+\boldsymbol{\nabla} \chi_{\rm C}({\bi r},t)
\label{transf}
\end{align}
satisfies the Poisson equation 
\begin{align}
\Delta \chi_{\rm C}({\bi r},t)=\frac{\partial\Phi_{\rm L}({\bi r},t)}{c\partial t},
\label{Poisson}
\end{align}
which follows from the second equality in (\ref{transf}), and  the conditions $\partial \Phi_{\rm L}/c\partial t+\boldsymbol{\nabla}\cdot{\bi A}_{\rm L} =0$ 
and $\boldsymbol{\nabla}\cdot{\bi A}_{\rm C}=0$ of the Lorenz and Coulomb gauges, respectively.

The Lorenz-gauge scalar potential $\Phi_{\rm L}({\bi r},t)$ of 
a  point charge $q$ moving with a constant velocity $ v\hat{\bi z}$ along the $z$-axis and  passing through the origin ${\bi r}=0$ at a time $t=0$ is given by 
\begin{align}
\Phi_{\rm L}({\bi r},t)=\frac{q}{\sqrt{s^2/\gamma^2+(z-v t)^2}},\quad \gamma=\frac{1}{\sqrt{1-v^2/c^2}},
\label{Phi_L}
\end{align}
the partial time derivative of which,
\begin{align}
\frac{\partial\Phi_{\rm L}({\bi r},t)}{c\partial t}=\frac{q\beta\gamma^3(z-vt)}{[s^2+\gamma^2(z-vt)^2]^{3/2}}, \quad \beta=\frac{v}{c},
\label{RHS}
\end{align}
decays as $1/r^2$ at $r\rightarrow \infty$. When the RHS of  
Poisson equation (\ref{Poisson}) is given by (\ref{RHS}), the equation cannot be solved  
using  the standard expansion of the inverse distance in spherical coordinates,
\begin{align}
\frac{1}{|{\bi r}-{\bi r}'|}=4\pi \sum_{l=0}^{\infty}\sum_{m=-l}^l\frac{1}{2l+1}\frac{r_<^l}{r_>^{l+1}}\,
Y_{lm}^*(\theta',\phi')Y_{lm}(\theta,\phi),
\end{align}
since the integrand of the requisite radial integral  then decays  only as 
$1/r$ at $r\rightarrow \infty$, and thus the integral representation of the solution  does not converge.

With a positive $a=\gamma$ and $b=vt$, and using that $1-\gamma^2= -\beta^2\gamma^2$,  the inverse distributional Laplacian (\ref{invLnondlt})
reads
\begin{align}
\bar{\Delta}^{-1} \frac{\beta^2\gamma^3(z-vt)}{[s^2+\gamma^2(z-vt)^2]^{3/2}}
={\rm arsinh}\frac{z-vt}{s}-{\rm arsinh}\frac{\gamma(z-vt)}{s},
\label{invLnondlt2}
\end{align}
which establishes that 
\begin{align}
\chi_{\rm C}({\bi r},t) =\frac{q}{\beta}\left[ {\rm arsinh}\frac{z-vt}{s}- {\rm arsinh}\frac{\gamma(z-vt)}{s}\right]
\label{chiC}
\end{align}
is the solution of the distributional  Poisson equation
\begin{align}
\bar{\Delta} \chi_{\rm C}({\bi r},t)=\frac{q\beta\gamma^3(z-vt)}{[s^2+\gamma^2(z-vt)^2]^{3/2}}.
\label{distPoisson}
\end{align}
A gauge transformation function equivalent to that of Eq.\,(\ref{chiC}) was obtained in \cite{VH}\footnote{See Eq.\,(13) in  \cite{VH}; it can be shown that  ${\rm arsinh}[(x-x_0)/s]$ there equals ${\rm arsinh}[\gamma(x-vt)/s]+{\rm arsinh}(\beta\gamma)$. } by applying  a formula of Jackson for $\chi_{\rm C}$ (\cite{Jack2}, Eq.\,(3.6)) to the case of a uniformly moving point charge.

\section{Concluding remarks}

The Poisson equation (\ref{distPoisson}) cannot be solved in spherical coordinates, but we solved it using an expansion of the inverse distance in cylindrical coordinates. The choice of the coordinates in which the Laplacian operator is expressed thus may determine whether a given Poisson equation is solvable or not.

Recently, the gauge transformation function $\chi_{\rm C}$ has been found for the case of  a point charge set suddenly from rest into uniform motion by integrating with respect to time the first equality in (\ref{transf}), where  $\Phi_{\rm C}$ and $\Phi_{\rm L}$ were the pertinent scalar potentials \cite{HV,Note}. The  gauge function obtained, which involves arsinh functions (see Eq.\,(43) in \cite{HV}), was confirmed  to satisfy the pertinent Poisson equation by evaluating its classical Laplacian. According to the findings of the present paper, when $t>0$, the  distributional Laplacian adds to that classical Laplacian a term\footnote{In \cite{HV} and \cite{Note}, the charge's motion is taken to be along the $x$-axis.}
\begin{align}
\frac{q}{\beta}\,[{\rm sgn}(x)-{\rm sgn}(x-vt)]\frac{\delta(s)}{s}\Theta(r-ct).
\label{extra}
\end{align}
It can be shown, however, that the inverse distributional Laplacian of this term vanishes, which makes the extra term (\ref{extra}) in this sense spurious.

\section*{Appendix}
\setcounter{equation}{0}
\renewcommand{\theequation}{A\arabic{equation}}
We outline here an informal proof of the validity of  Eq.\,(\ref{limDelta}) for any well-behaved test function $f(s)$. The `epsilon-regularized' Laplacian in (\ref{limDelta}) evaluates to 
\begin{align}
\Delta\ln[a(z-b)+\sqrt{s^2+a^2(z-b)^2+\epsilon^2}]=\frac{(1-a^2)A}{(s^2+A^2+\epsilon^2)^{3/2}}
+\epsilon^2(B+C),
\label{DeltaLnEps}
\end{align}
where
\begin{align}
A&=a(z-b), \label{A}\\
B&=\frac{2 }{(s^2+A^2+\epsilon^2)(A+\sqrt{s^2+A^2+\epsilon^2})^2}, \label{B}\\
C&=\frac{A}{(s^2+A^2+\epsilon^2)^{3/2}(A+\sqrt{s^2+A^2{+}\epsilon^2})^2} \label{C}.
\end{align}

The limit $\epsilon\rightarrow 0$ of the first term on the RHS of (\ref{DeltaLnEps}) is manifestly finite, and so the  first term on the RHS of  (\ref{limDelta}) is  accounted for immediately.
Expanding the test function $f(s)$ in a Taylor series around $s=0$, assumed to converge in an interval $0\le s<S$, we can write
\begin{align}
\int_0^{\infty} s\,\rmd s\,f(s)\epsilon^2(B +C)= \sum_{n=0}^{\infty}\frac{f^{(n)}(0)}{n!}
\epsilon^2\int_0^S \rmd s\,s^{n+1}(B +C)
 +\epsilon^2\int_S^{\infty} s\,\rmd s\,f(s)(B +C).
\label{intBC}
\end{align}
Here, the limit  $\epsilon\rightarrow 0$ of the 2nd term on the RHS  vanishes since the integral it involves  converges to a finite value when $f(s)$  is a well-behaved test function.

Let us  now investigate  the limit  
\begin{align}
\lim_{\epsilon\rightarrow 0} \left[\epsilon^2\int_0^S \rmd s\,s^{n+1} B\right]
&=\lim_{\epsilon\rightarrow 0}\left[2\epsilon^2\int_0^S 
\frac{ \rmd s\,s^{n+1}}{(s^2+A^2+\epsilon^2)(A+\sqrt{s^2+A^2+\epsilon^2})^2}\right]\nonumber \\
&= \lim_{\epsilon\rightarrow 0}\left[2\epsilon^2\int_{\sqrt{A^2+\epsilon^2}}^{\sqrt{S^2+A^2+\epsilon^2}}\rmd u\,\frac{(u^2-A^2-\epsilon^2)^{n/2}}{u(u+A)^2}\right].
\label{limB}
\end{align}
Here, the integration variable $s$ was transformed to
 $u=\sqrt{s^2+A^2+\epsilon^2}$, so that $\rmd s=u\,\rmd u/\sqrt{u^2-A^2-\epsilon^2}$.
For $n=0$, the RHS of  (\ref{limB})  equals
\begin{align}
\lim_{\epsilon\rightarrow 0}2\epsilon^2\left[\frac{1}{A(u+A)}+\frac{1}{A^2}\ln\frac{u}{u+A}\right]_{u=\sqrt{A^2+\epsilon^2}}^{\sqrt{S^2+A^2+\epsilon^2}}.
\label{nzero}
\end{align}
This limit vanishes when $A$ is positive, but when $A$ is negative, it equals 4, which is contributed by the lower limit of the first term in the brackets,
\begin{align}
 -\lim_{\epsilon\rightarrow 0}
\frac{2\epsilon^2}{A(\sqrt{A^2+\epsilon^2}+A)}&=-\lim_{\epsilon\rightarrow 0}\frac{4\sqrt{A^2+\epsilon^2}}{A}\nonumber \\
&= -4\,\frac{|A|}{A}=4,
\label{l'Hop}
\end{align}
where l'Hopital's rule is used.
For $n>0$, the integral in (\ref{limB}) can be evaluated in terms of the Appell function $F_1$ and the hypergeometric function $_2F_1$. As $\epsilon\rightarrow 0$, the arguments of both functions approach $\infty$; their 
asymptotic behavior at large arguments can be seen to be  as $\epsilon^n$ (\cite{Ferr}; \cite{AS}, item 15.7.1), and so their limits $\epsilon\rightarrow 0$ vanish.

The limit (\ref{limB}) is thus
\begin{align}
\lim_{\epsilon\rightarrow 0}\left[ \epsilon^2\int_0^S \rmd s\,s^{n+1} B\right]=2[1-{\rm sgn}(A)]\delta_{n0}.
\label{limitB}
\end{align}
An investigation similar to that in the preceding  paragraph yields 
\begin{align}
\lim_{\epsilon\rightarrow 0} \left[\epsilon^2\int_0^S \rmd s\,s^{n+1} C\right]=-[1-{\rm sgn}(A)]\delta_{n0}.
\label{limitC}
\end{align}
Using  (\ref{limitB}), (\ref{limitC}) and the other results of this Appendix, we have 
\begin{align}
&\lim_{\epsilon\rightarrow 0}\int_0^{\infty} s\,\rmd sf(s)\,
\Delta\ln[a(z-b)+\sqrt{s^2+a^2(z-b)^2+\epsilon^2}] \nonumber \\
&\quad=(1-a^2)A\int_0^{\infty}\frac{s\,\rmd s f(s)}{(s^2+A^2)^{3/2}}+\lim_{\epsilon\rightarrow 0}
\sum_{n=0}^{\infty}\frac{f^{(n)}(0)}{n!}\,\epsilon^2\int_0^S \rmd s\,s^{n+1}(B+ C)
\nonumber \\
&\quad=a(1-a^2)(z-b)\int_0^{\infty}\frac{s\,\rmd s f(s)}{(s^2+a^2(z-b)^2)^{3/2}}+
\big[1-{\rm sgn}[a(z-b)]\big]f(0),
\end{align}
which accomplishes our task.

\end{document}